# A new BAT for Acyclic Multistate Information Network Reliability Evaluation

Wei-Chang Yeh
Integration and Collaboration Laboratory
Department of Industrial Engineering and Engineering Management
National Tsing Hua University
yeh@ieee.org

**Abstract**: The acyclic multistate information network (AMIN), which is a kind of MIN that does not require the conservation law of flow, plays an important role nowadays because many modern network structures present AMIN as the construction such as social networks, local area networks (LANs), 4G/5G networks, etc. To effectively evaluate the network reliability of AMIN, which indicates the reliable operation of the network, showing a major and primary metrics for determining the performance and quality of the overall network. The network reliability, which has been shown a NP-hard, has been successfully resolved and approached by the universal generation function method (UGFM). However, the UGFM can only solve small-scale problems due to the overflow in computer memory. To overcome the memory obstacle, an improved and enhanced binary-addition vectors tree algorithm (BAT) is proposed to effectively evaluate and analyze the reliability of AMIN. The performance of the proposed BAT is validated on examples.

**Keywords**: Acyclic multistate information network (AMIN); Network reliability; Universal Generation Function Method (UGFM); Binary-Addition Tree Algorithm (BAT).

## 1. INTRODUCTION

The multistate networks, which compose of multistate components and own multiple distinct levels of capacity [1], play a very important position and possess a very powerful and wide range of applications in practical applications, especially nowadays with the rapid development and dependence of network technology. Therefore, there are many applications of multistate networks research in various fields, such as wireless sensor networks (WSN) [2, 3], social networks [4], commodity distribution [5], power systems [6], satellite systems [7], transportation systems [8], and manufacturing systems [9]. To guarantee the reliable operation, the design and operation of system must possess the indicator of performance measurement. Reliability, which shows the probability of successful system operation, is one of the major and popular indicators used to evaluate and measure the performance of multistate networks [1-29].

Network reliability plays a significant role in modern society. Thus, how to effectively evaluate the network reliability plays an important research work in the modern era. Network reliability is classified into two categories: the multistate flow network (MFN) meets the conservation law of flow [1-3, 5-9, 13-29] but the multistate information network (MIN) does not meet the conservation law of flow [28, 30-33] according to the flow or signal whether matches the conservation law of flow, which indicates flow in equals flow out.

It is necessary to enhance the MIN research because most of the network reliability research are focused on MFN [1-3, 5-9, 13-29] and a few are on MIN [28, 30-33]. The MIN usually shows the architecture of a tree-structured multistate network. In addition, many modern network structures present MIN as the construction, such as computer networks and cellular phone networks [28, 30-33].

Each node in MIN possesses multiple different states defined according to a group of nodes, which straightly receive information, without requiring to meet the conservation law of flow [28, 30-33]. It contains a source node that is just responsible for transmitting information to other nodes, several sink nodes that are just responsible for receiving information, and a series of intermediate nodes, which are not source node and not sink nodes, that are responsible for transmitting the received information to non-source nodes. Therefore, the information, which received by the source node, is transmitted the received information to the sink nodes by the intermediate nodes via the edges connecting between nodes.

The computation of network reliability was shown to be NP-hard by ?. in 1980 [34]. Numerous scholars invest in different algorithms to approach and solve various problems of network reliability. Most of the network reliability problems of MFN are approached and solved by all kinds of methods modified from minimal path-based (MPs) or minimal cut-based (MCs) algorithms, for example, based on MPs including an enumeration of shortest technique [2], a sequential decomposition approach [5], a simple algorithm for multiple commodity of unreliable network [13], a sum-of-disjoint products approach [14], an approximation method with considering the network diameter [15], an ordering heuristic approach [21], an addition-based technique [23], as well as a mimics natural organisms method [26] and based on MCs containing an intuitive algorithm [1], a lower bound level of flow approach [19], as well as an ordering heuristic approach for two-terminal network [21]. Nevertheless, most of the network reliability problems of MIN are analyzed and resolved by the derived algorithms from the universal generating function method (UGFM) that was first developed by Levitin in 2004 [35], and enhanced subsequently by Yeh [31-33], such as a novel label UGFM [31], a straightforward UGFM algorithm for the network of multiple sources and multiple targets [32], and a convolution UGFM [33].

Since its initial introduction [16], the UGFM has proven to be very effective for evaluating the exact symbolic reliability of AMIN [16]-[24], and it does not demand rigorous computational effort. Further developments and applications of UGFM were presented in [20]-[24], and detailed description can be found in [20]-[25], which summarize the recent achievements in the field of network reliability estimation based on UGFM. Also, a fuzzy based UGF was proposed to solve the problem for which the performance rates and probabilities of states cannot be exactly determined in the recent research [25].

The main idea of UGFM is to extend state sub-vectors of which the number coordinates is less than that of state vector parts to state vectors in a heuristic way. Hence, UGFM needs to have enough computer memory to store sub-vectors and the number of sub-vectors is increased by the network size due to the natural characteristic of #P-hard problems. Thus, UGFM is failed in high dense networks which are very common and practical now. To overcome the computer memory obstacle of UGFM, a new algorithm based on binary-addition tree algorithm (BAT) is proposed.

The (arc-based) BAT first proposed by Yeh recently [] is developed based on easy-understanding binary-addition process. The BAT is easy to modify, code, and apply. Moreover, it outperforms depth-first-search (DFS) based algorithms [21, 22] and breadth-first-search (BFS) based algorithms from experiments in the required computer memory and computational efficiency [20]. Therefore, a novel node-based BAT is proposed for the AMIN reliability problem in this paper.

The rest of this study is organized as follows. The required acronym, notations, nomenclature, and assumptions are listed in Section 2. The overview of the MIN, UGFM, and BAT are provided in Section 3. The details of the new node-based BAT for the AMIN reliability problem are proposed in Section 4 together with the provided state label based on another BAT, the related pseudo code, the discussion of the time complexity and computer memory, and an example to illustrate how to implement the proposed new BAT. In Section 5, a comparison between the proposed BAT and UGFM is provided to validate the performance of the new BAT and the shortcoming of the UGFM in managing the computer memory. Concluding remarks and summarizes the discussion are given in Section 6.

## 2. ACRONYM, NOTATIONS, NOMENCLATURE, AND ASSUMPTIONS

### 2.1  ACRONYM

MIN    Multistate information network

MFN    Multistate flow network

AMIN    Acyclic MIN

LAN    Local area networks

WSN    Wireless sensor networks

MP    Minimal path

MC    Minimal cut

UGF    Universal generating function

UGFM    UGF method

BAT    Binary-addition tree algorithm

DFS    Depth-first-search

BFS    Breadth -first-search

### 2.2  NOTATIONS

$|\bullet|$    number of elements in set $\bullet$.

$E$   set of directed arcs.

$e_{i,j}$   $e_{i,j} \in E$ such that information can be transmitted directly from nodes $i$ to $j$.

$m$   number of arcs, i.e., $|E| = m$.

$V$   set of nodes $V=\{1, 2, \ldots, n\}$, node 1 is the source node, and $i < j$ if arc $e_{i,j} \in E$ for all nodes $i$ and $j$ in $V$.

$G(V, E)$   a connected AMIN with $V$ and $E$, e.g. the network in Fig. 1 is an AMIN.

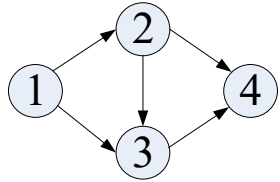

Fig. 1. An AMIN example.

$V_i$   $V_i = \{ j \in V \mid$ for all $e_{i,j} \in E\}$ is an ordered mode subset for all $i \in V$ and all nodes in $V_i$ are arranged in the decreasing order of labels, e.g., $V_1 =\{2, 3\}$ and $V_2 =\{3, 5\}$ in Fig. 1.

$\mathrm{Deg}(i)$   $\mathrm{Deg}(i) = |V_i|$ is the degree of node $i \in V$, e.g., $\mathrm{Deg}(1) = \mathrm{Deg}(2) = 2$ in Fig. 1.

$V_i(u)$   $V_i(u)$ is the position of node $u$ in $V_i$, e.g., $V_1(2) =1$ and $V_1(3) =2$ in Fig. 1.

$p_{i,I}$   $p_{i,I} = \mathrm{Pr}\{$only these nodes in $I$ receiving information directly from $i\}$, where $i \in V$ and $I \subseteq V_i$.

$O(f(n))$   $O(f(n)) = \lim\limits_{n\to\infty} c \cdot f(n)$   for some constant $c$.

$u(i)$   the node-UGF of node $i = 1, 2, \ldots, |V|-1$.

$U(i)$   the subnet-UGF of node $i = 1, 2, \ldots, |V|-1$ and $U(1) = u(1)$.

$\otimes$   the universal generating operator for UGFM.

$R_{1:J}$   the probability of the target node receiving information from node 1.

$S_{i,j}$   the $j$th state of node $i$, i.e., the $j$th node subset in $V_i$, and the integer $j$ is the state label denoting which state is discussed now.

$S_{i,j}(u)$   $S_{i,j}(u) = 0$ if $u \notin S_{i,j}$, and $S_{i,j}(u) = 1$ if $u \in S_{i,j}$.

$B(S_{i,j})$   a binary vector corresponding to $S_{i,j}$ such that its $k$th coordinate is equal to $S_{i,j}(u)$ for all $V_i(u) = k$.

$B_{i,j}$   A binary vector corresponding to the $j$th state of node $i$, $B_{i,j}(u)$ be the value of the coordinated represented node $u$, and $P(B_{i,j}(u))$ be the position of the coordinated represented node $u$.

## 2.3 NOMENCLATURE

| | |
|---|---|
| Acyclic network | A directed network contains no directed cycle. |
| Target set | a non-empty node subset and only its node can receive information. |
| Network reliability $R_J$ | the successful probability in transmitting the formation from node 1 to specific nodes in target set. |
| Flow conservation law | the total in flow is equal to the total out flow for all nodes excepted node 1 and target nodes. |
| Operator $\otimes$ | the operator to unify node-UGFs with subnet-UGFs. |
| node-UGF | the basic UGF of UGFMs. |
| subnet-UGF | the UGF unified using the operator $\otimes$ by node-UGFs. |
| NP-hard | a problem is NP-hard if and only if there is an NP-complete problem can be Turing-reducible to such problem in polynomial time [14]. |

State - One of the subsets of the combination of reachable nodes, e.g. 2:$\varnothing$, 2:{3}, 2:{5}, and 2:{3,5} are all states of node 2 in Fig. 1.

Independent term - The individual product in the UGF, e.g. $p_{1:\{2\}}z^{\{2\}}$, $p_{1:\{3\}}z^{\{3\}}$ and $p_{1:\{2,3\}}z^{\{2,3\}}$ are independent terms in $U(1)=u(1)=p_{1:\{2\}}z^{\{2\}}+p_{1:\{3\}}z^{\{3\}}+p_{1:\{2,3\}}z^{\{2,3\}}$. Basically, each term in $U(i)=$ $\sum_{\substack{I\subseteq\Theta_i \\ \Theta_i\neq\varnothing}}\pi_{i,I}z^I$ represents the probability of a directed acyclic network connected to the source node, and $\Theta_i$ ($\neq\varnothing$) such that no redundant arcs/nodes are included.

## 2.4 ASSUMPTIONS

The following assumptions are satisfied by AMINs:

1. All nodes and arcs are perfectly reliable.

2. $\sum_{I\subseteq V_i} p_{i,I} = 1$ and $p_{i,I}$ is a $s$-independent random variable for all node $i \in 1, 2, \ldots, (|V| - 1)$ and $I \subseteq V_i$.

3. No undirected arcs and directed cycles exist.

## 3. OVERVIEW OF MIN, UGFM AND BAT

The proposed new node-based BAT is based on the traditional BAT to solve the AMIN reliability problems and compare to the UGFM which is the main algorithm for the problem. Therefore, before considering the proposed BAT, an overall review of the AMIN, UGFM, and BAT are presented in this section.

### 3.1 AMIN and MIN

A multistate information network (MIN) generalizes the tree-structured multistate-state system without satisfying the flow conservation law. The acyclic multistate information network (AMIN), which refers to the MIN where the information transmission within the network is not allowed to have loop that indicates the information transmitted from any node will not be transmitted back via any other node, is another typical class of MIN. Hence, the labels of nodes must be arranged in a way such that $e_{i,j} \notin E$ if $i < j$ in AMIN reliability problems.

The AMIN/MIN is more care about where the information from and to rather the amount of flows in the MFN. A state of a node $i$ is a node subset in $V(i)$ if $i$ receives information, and these nodes in a state all receives information sent from node $i$ for all node $i = 1, 2, \ldots, n-1$. Hence, the state of node $i$ is defined as the node subset that received information from node $i$ and whether node $i$ receives the information in AMIN/MIN.

The current known methods for the AMIN reliability problems are developed from the UGFM to find all feasible state vectors of which node 1 and $n$ are connected without redundant nodes in their corresponding subgraphs. The effectiveness of the UGFM in solving the AMIN reliability, which also belongs NP-hard, has been shown by various research including evaluating the reliability of AMIN [28, 33, 36-39], assessing the reliability of AMFN [39, 40], and determining the reliability of acyclic binary flow network (ABFN) [41].

## 3.2 UGFM

The UGFM is an extension of the Boolean models for the cases with multi-values [42, 43]. There are node-UGF and subnet-UGF, two types of UGFs, in the traditional UGFM [31, 32]. The node-UGF $u(i)$ and subset-UGF $U(i)$ of node $i$ for $i = 1, 2, \ldots, n-1$ are defined as follows [31, 32]:

$$u(i) = \sum_{J \subseteq V_i} p_{i:J} z^J , \tag{1}$$

$$U(i) = U(i-1) \otimes u(i), \tag{2}$$

where the subscript of $z$ denotes the node subset received the information from $i$, notation $z$ is just to separate $p_{i:J}$ and the subscript of $z$, and operator $\otimes$ is similar to multiplication to unify the node-UGFs in a recursive expression [31-33].

The UGFM must follow the corresponding order of node labels in finding subset-UGF to examine all possible states of each node comprehensively. Compared to the binary-addition tree algorithm (BAT), the UGFM is still too complicated in terms of number of symbols and functions. Also, the number of sub-vectors, call terms in UGFM, in $U(i)$ is increased exponentially with $i$ and the AMIN size because of the NP-Hard characteristic.

The pseudo code for the traditional UGFM is described as follows.

**STEP U0.** Find all node-UGFs except the target node:

$$u(i) = \begin{cases} \displaystyle\sum_{J \subseteq V_i \text{ and } J \neq \varnothing} p_{i:J} z^J & i = 1 \\ \displaystyle\sum_{J \subseteq V_i} p_{i:J} z^J & 1 < i < n \end{cases} . \tag{34}$$

**STEP U1.** Let $U(1) = u(1)$ and $i = 2$.

**STEP U2.** Calculate and simplify $U(i) = U(i-1) \otimes u(i)$.

**STEP U3.** If $i < n$, go to STEP U2.

**STEP U4.** $R_{\{n\}} = r_{\{n\}}$ obtained from $U(i-1)$ is the AMIN reliability from nodes 1 to $n$.

In the traditional UGFM, the number of terms in $u(i)$ is $O(2^{\text{Deg}(i)})$ in STEP U1. The main time complexity is made from STEP U2 and its time complexity for finding $U(i)$ can be expressed of $U(i-1) \cdot u(i) = O(2^{\varpi(i-1)}) \cdot O(2^{\text{Deg}(i)}) = O(2^{\varpi(i)})$, where $\varpi(i) = \sum_{i=1}^{i} \text{Deg}(i)$, i.e., the time complexity is $O(2^{\varpi(n)}) = O(2^{|E|})$ in STEP 2. Besides, it takes $O(|V|)$ to verify whether a term in $U(n-1)$ is a feasible state vector and calculate $R(X)$ in STEP U4. Hence, the total time complexity is $O(|V|2^{|E|})$. Moreover, the computer memory to store $U(i)$ is also $O(2^{\varpi(i)})$, i.e., $O(2^{|E|})$ for $i = n - 1$.

For example, consider the AMIN presented in Fig. 1, all node-UGFs from STEP U0 are listed below:

$$u(1) \quad = p_{1,\varnothing} z^{\varnothing} + p_{1:\{2\}} z^{\{2\}} + p_{1:\{3\}} z^{\{3\}} + p_{1:\{2,3\}} z^{\{2,3\}} \tag{31}$$

$$= p_{1:\{2\}} z^{\{2\}} + p_{1:\{3\}} z^{\{3\}} + p_{1:\{2,3\}} z^{\{2,3\}} \tag{32}$$

$$u(2) \quad = p_{2,\varnothing} z^{\varnothing} + p_{2:\{3\}} z^{\{3\}} + p_{2:\{4\}} z^{\{4\}} + p_{2:\{3,4\}} z^{\{3,4\}} \tag{33}$$

$$u(3) \quad = p_{3,\varnothing} z^{\varnothing} + p_{3:\{4\}} z^{\{4\}}. \tag{34}$$

Based on STEPs U1–U3, all subnet-UGFs are obtained as follows.

$$U(1) \quad = u(1)$$
$$= p_{1:\{2\}} z^{\{2\}} + p_{1:\{3\}} z^{\{3\}} + p_{1:\{2,3\}} z^{\{2,3\}}. \tag{36}$$

$$U(2) \quad = U(1) \otimes u(2)$$
$$= (p_{1:\{2\}} z^{\{2\}} + p_{1:\{3\}} z^{\{3\}} + p_{1:\{2,3\}} z^{\{2,3\}}) \otimes (p_{2,\varnothing} z^{\varnothing} + p_{2:\{3\}} z^{\{3\}} + p_{2:\{4\}} z^{\{4\}} + p_{2:\{3,4\}} z^{\{3,4\}})$$
$$= (p_{1:\{3\}} + p_{1:\{2,3\}} p_{2,\varnothing} + p_{1:\{2,3\}} p_{2:\{3\}} + p_{1:\{2\}} p_{2:\{3\}}) z^{\{3\}} + p_{1:\{2\}} p_{2:\{4\}} z^{\{4\}} +$$
$$(p_{1:\{2,3\}} p_{2:\{4\}} + p_{1:\{2,3\}} p_{2:\{3,4\}} + p_{1:\{2\}} p_{2:\{3,4\}}) z^{\{3,4\}} \tag{40}$$

$$U(3) \quad = U(2) \otimes u(3)$$
$$= U(2) \otimes (p_{3,\varnothing} z^{\varnothing} + p_{3:\{4\}} z^{\{4\}})$$
$$= [(p_{1:\{3\}} + p_{1:\{2,3\}} p_{2,\varnothing} + p_{1:\{2,3\}} p_{2:\{3\}} + p_{1:\{2\}} p_{2:\{3\}}) z^{\{3\}} +$$
$$(p_{1:\{2,3\}} p_{2:\{4\}} + p_{1:\{2,3\}} p_{2:\{3,4\}} + p_{1:\{2\}} p_{2:\{3,4\}}) z^{\{3,4\}}] \otimes (p_{3,\varnothing} z^{\varnothing} + p_{3:\{4\}} z^{\{4\}})$$
$$= (p_{1:\{3\}} + p_{1:\{2,3\}} p_{2,\varnothing} + p_{1:\{2,3\}} p_{2:\{3\}} + p_{1:\{2\}} p_{2:\{3\}}) p_{3:\{4\}} z^{\{4\}} +$$

$$(p_{1:\{2,3\}}p_{2:\{4\}}+p_{1:\{2,3\}}p_{2:\{3,4\}}+p_{1:\{2\}}p_{2:\{3,4\}})p_{3,\varnothing}z^{\{4\}} +$$

$$(p_{1:\{2,3\}}p_{2:\{4\}}+p_{1:\{2,3\}}p_{2:\{3,4\}}+p_{1:\{2\}}p_{2:\{3,4\}})p_{3:\{4\}}z^{\{4,4\}}+ p_{1:\{2\}}p_{2:\{4\}}z^{\{4\}} \quad (47)$$

Hence, we have eleven state vectors (terms) as follows:

$$R_{1:\{4\}} = (p_{1:\{3\}}+p_{1:\{2,3\}}p_{2,\varnothing}+p_{1:\{2,3\}}p_{2:\{3\}}+p_{1:\{2\}}p_{2:\{3\}})p_{3:\{4\}} +$$

$$(p_{1:\{2,3\}}p_{2:\{4\}}+p_{1:\{2,3\}}p_{2:\{3,4\}}+p_{1:\{2\}}p_{2:\{3,4\}})p_{3,\varnothing} +$$

$$(p_{1:\{2,3\}}p_{2:\{4\}}+p_{1:\{2,3\}}p_{2:\{3,4\}}+p_{1:\{2\}}p_{2:\{3,4\}})p_{3:\{4\}} +$$

$$p_{1:\{2\}}p_{2:\{4\}} \quad (47)$$

$$= p_{1:\{3\}}p_{3:\{4\}} + p_{1:\{2,3\}}p_{2,\varnothing}p_{3:\{4\}} + p_{1:\{2,3\}}p_{2:\{3\}}p_{3:\{4\}} + p_{1:\{2\}}p_{2:\{3\}}p_{3:\{4\}} +$$

$$p_{1:\{2,3\}}p_{2:\{4\}}p_{3,\varnothing} + p_{1:\{2,3\}}p_{2:\{3,4\}}\ p_{3,\varnothing} + p_{1:\{2\}}p_{2:\{3,4\}}p_{3,\varnothing} +$$

$$p_{1:\{2,3\}}p_{2:\{4\}}p_{3:\{4\}} + p_{1:\{2,3\}}p_{2:\{3,4\}}p_{3:\{4\}} + p_{1:\{2\}}p_{2:\{3,4\}}p_{3:\{4\}} +$$

$$p_{1:\{2\}}p_{2:\{4\}}. \quad (47)$$

The corresponding subgraphs of these eleven terms in $R_{1:\{4\}}$ are listed below:

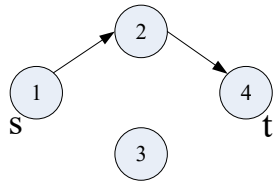
(a) $p_{1:\{2\}}p_{2:\{4\}}$

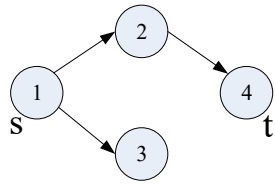
(b) $p_{1:\{2,3\}}p_{2:\{4\}}p_{3,\varnothing}$

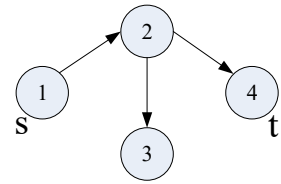
(c) $p_{1:\{2\}}p_{2:\{3,4\}}p_{3,\varnothing}$

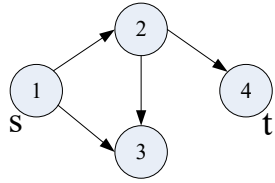
(d) $p_{1:\{2,3\}}p_{2:\{3,4\}}p_{3,\varnothing}$

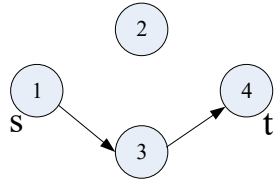
(e) $p_{1:\{3\}}p_{3:\{4\}}$

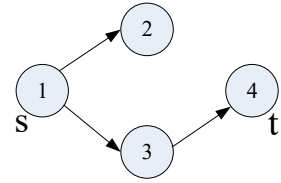
(f) $p_{1:\{2,3\}}p_{2,\varnothing}p_{3:\{4\}}$

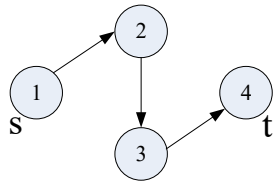
(g) $p_{1:\{2\}}p_{2:\{3\}}p_{3:\{4\}}$

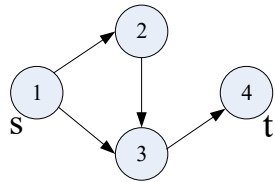
(h) $p_{1:\{2,3\}}p_{2:\{3,4\}}p_{3:\{4\}}$

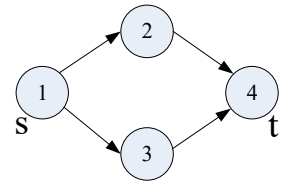
(i) $p_{1:\{2,3\}}p_{2:\{4\}}p_{3:\{4\}}$

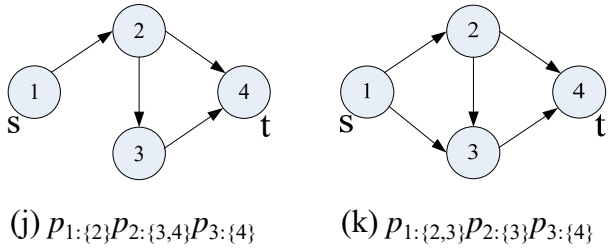

(j) $p_{1:\{2\}}p_{2:\{3,4\}}p_{3:\{4\}}$      (k) $p_{1:\{2,3\}}p_{2:\{3\}}p_{3:\{4\}}$

**Figure ?.** The subgraphs of related feasible state vectors.

Note that before $U(i+1)$ is formed, each sub-vector obtained in subnet-UGF $U(i)$, e.g., $p_{1:\{2\}}$ in $U(1)$ and $p_{1:\{2,3\}}p_{2:\{4\}}$ in $U(2)$, cannot be discarded and needed to spent extra space store for $i = 1, 2, \ldots, n-1$.

### 3.3 BAT

The original BAT is a simple but efficient algorithm in generating all the possible binary-state vectors of which each coordinate value is either 0 or 1. There is always an array, i.e., the state vector $X$, in the BAT. The $X$ is updated in each loop and the updated $X$ is processed under some predefined goals, i.e., calculates the state probability, verifies whether it is a feasible vector, tests nodes 1 and $|V| = n$ are still connected, counts the number of nodes or arcs in the subgraph corresponding to $X$, etc. After that such $X$ is updated again and again without needing to save all information of the precedence of current $X$.

The overall pseudo-code of BAT is listed in the following [20].

**<u>BAT []</u>**

**Input:**      A network $G(V, E)$ with two states, either 0 or 1, to each arc and $V = \{1, 2, \ldots, n\}$.

**Output:**     All possible non-duplicate arc-based state vectors.

**STEP B0.** Let $X = 0$ with $|E|$ coordinates, $i = 1$, and SUM $= 0$.

**STEP B1.** If $X(i) = 0$, let $X(i) = 1$, SUM $=$ SUM $+ 1$, and go to STEP B4.

**STEP B2.** Let $X(i) = 0$ and SUM $=$ SUM $- 1$.

**STEP B3.** If $i < |E|$, let $i = i + 1$ and return to STEP B1.

**STEP B4.** If SUM $= |E|$, halt; otherwise, let $i = 1$ and return to STEP B1.

Started from zero vector as shown in STEP B0, the binary number of $X$ is updated by adding one as listed in STEPs B1-B3 and proceeds some functions to $X$ in STEP B4. Repeat the above process until all the coordinates are 1, i.e., SUM = $m$ as provided in STEP B4.

Rewrite the state vector $X = (b_1, b_2, \ldots, b_5)$ to $b_1b_2\ldots b_{|E|}$, where $b_i$ is either 0 or 1 for $i = 1, 2, \ldots, |E|$, e.g., the binary numbers of (0, 1, 1, 1, 1) and (0, 0, 0, 0, 0) are 01111 and 00000, respectively. In Fig. 1 where $|E| = 5$, the first new five vectors updated from $X = (0, 0, 0, 0, 0)$ are showed below:

$$00000 + 1 = 00001, \tag{5}$$

$$00001 + 1 = 00010, \tag{6}$$

$$00001 + 1 = 00011, \tag{7}$$

$$00011 + 1 = 00100. \tag{8}$$

$$00101 + 1 = 00110. \tag{8}$$

Correspondingly, we have $2^5 = 32$ different vectors in total because the each coordinate is either 0 or 1 and there are five coordinates in $X$. All non-duplicate vectors can be found using the BAT as shown in Table 3.

**Table 3.** all vectors obtained from BAT [20].

| $i$ | $X$ | $i$ | $X$ |
|---|---|---|---|
| 1 | (0, 0, 0, 0, 0) | 17 | (1, 0, 0, 0, 0) |
| 2 | (0, 0, 0, 0, 1) | 18 | (1, 0, 0, 0, 1) |
| 3 | (0, 0, 0, 1, 0) | 19 | (1, 0, 0, 1, 0) |
| 4 | (0, 0, 0, 1, 1) | 20 | (1, 0, 0, 1, 1) |
| 5 | (0, 0, 1, 0, 0) | 21 | (1, 0, 1, 0, 0) |
| 6 | (0, 0, 1, 0, 1) | 22 | (1, 0, 1, 0, 1) |
| 7 | (0, 0, 1, 1, 0) | 23 | (1, 0, 1, 1, 0) |
| 8 | (0, 0, 1, 1, 1) | 24 | (1, 0, 1, 1, 1) |
| 9 | (0, 1, 0, 0, 0) | 25 | (1, 1, 0, 0, 0) |
| 10 | (0, 1, 0, 0, 1) | 26 | (1, 1, 0, 0, 1) |
| 11 | (0, 1, 0, 1, 0) | 27 | (1, 1, 0, 1, 0) |
| 12 | (0, 1, 0, 1, 1) | 28 | (1, 1, 0, 1, 1) |
| 13 | (0, 1, 1, 0, 0) | 29 | (1, 1, 1, 0, 0) |
| 14 | (0, 1, 1, 0, 1) | 30 | (1, 1, 1, 0, 1) |
| 15 | (0, 1, 1, 1, 0) | 31 | (1, 1, 1, 1, 0) |
| 16 | (0, 1, 1, 1, 1) | 32 | (1, 1, 1, 1, 1) |

# 4. PROPOSED NEW NODE-BASED BAT

The major components of the proposed new node-based BAT together with the pseudo code explains the way to implement the node-based BAT and an example demonstrates how to perform the proposed node-based BAT are all provided in this section.

## 4.1 Node State Label and Probability

The ordered node subset $V_i = \{ j \in V \mid$ for all $e_{i,j} \in E \}$ is a special set such that $V_i(u) < V_i(v)$ if $u < v$ for all $u, v \in V_i$, i.e., node $i$ is adjacent to all nodes in $V_i$ of which all nodes are arranged in the decreasing order of node labels, where $V_i(u)$ is the position of $u$ in $V_i$. For example, because $V_1 = \{2, 3\}$ in Fig. 1, $V_i(2) = 1 < V_i(3) = 2$.

Any subset of $V_i$ is a node state of node $i$. Let $S_{i,j}$ be the $j$th node state of node $i$, i.e., the $j$th node subset in $V_i$. Node $i$ maybe received no information, thus, we define $S_{i,j} = $ "Z" for such node $i$. Let $S_{i,j}(u) = 1$ if $u \in S_{i,j}$. For example, any subset in $V_1 = \{2, 3\}$ is a state, i.e., $S_{1,1} = \varnothing$ and $S_{1,1}(2) = S_{1,1}(3) = 0$; $S_{1,2} = \{2\}$, $S_{1,2}(2) = 1$, and $S_{1,3}(3) = 0$; $S_{1,3} = \{3\}$, $S_{1,3}(2) = 0$, and $S_{1,3}(3) = 1$; $S_{1,4} = \{2, 3\}$, and $S_{1,4}(2) = S_{1,4}(3) = 1$. Note that there are $2^{|V_i|}$ states for any node $i$ received information.

To implement the proposed new node-based BAT easily, each state is labeled to be a binary vector. Let $B(S_{i,j})$ be the binary vector corresponding to $S_{i,j}$ such that the $k$th coordinate of $B(S_{i,j})$ is $S_{i,j}(u)$ for all $V_i(u) = k$ if node $i$ receives information. Also, we have the decimal value of the binary vector $B_{i,j}(u)$ is equal to $j$, i.e., $B(S_{i,j}(k_1)) + 2 \cdot B(S_{i,j}(k_2)) + 2^2 \cdot B(S_{i,j}(k_3)) + \ldots + 2^{(\kappa-1)} \cdot B(S_{i,j}(k_\kappa)) = j$ and $\kappa = |V_i|$.

For example, in Fig. 1, $B(S_{1,1}) = (0, 0)$ and the decimal of "00" is 0 because $S_{1,1} = \varnothing$; $B_{1,2} = (0, 1)$ and the decimal number of "01" is 2 because $S_{1,2} = \{2\}$; $B_{1,3} = (1, 0)$ and the decimal number of "10" is 3 because $S_{1,3} = \{3\}$; $B_{1,4} = (1, 1)$ and the decimal number of "11" is 3 because $S_{1,4} = \{2, 3\}$. In the same way, we have all node state labels, their corresponding binary vectors, and the related probabilities are provided in Table ?.

**Table 1.** Example of $S_{i,j}$, $B(S_{i,j})$, and $\Pr(S_{i,j})$.

| $i$ | $j$ | $V_i$ | $S_{i,j}$ | $B(S_{i,j})$ | $\Pr(S_{i,j})$ |
| --- | --- | --- | --- | --- | --- |

| 1 | 0 | $\{2, 3\}$ | Z | | $p_{1,0} = 1$ |
|---|---|---|---|---|---|
| | 1 | | $\varnothing$ | $(0, 0)$ | $p_{1:\varnothing} = p_{1,1}$ |
| | 2 | | $\{2\}$ | $(0, 1)$ | $p_{1:\{2\}} = p_{1,2}$ |
| | 3 | | $\{3\}$ | $(1, 0)$ | $p_{1:\{3\}} = p_{1,3}$ |
| | 4 | | $\{2, 3\}$ | $(1, 1)$ | $p_{1:\{2, 3\}} = p_{1,4}$ |
| 2 | 0 | $\{3, 4\}$ | Z | | $p_{2,0} = 1$ |
| | 1 | | $\varnothing$ | $(0, 0)$ | $p_{2:\varnothing} = p_{2,1}$ |
| | 2 | | $\{3\}$ | $(0, 1)$ | $p_{2:\{3\}} = p_{2,2}$ |
| | 3 | | $\{5\}$ | $(1, 0)$ | $p_{2:\{5\}} = p_{2,3}$ |
| | 4 | | $\{3, 5\}$ | $(1, 1)$ | $p_{2:\{3, 5\}} = p_{2,4}$ |
| 3 | 0 | $\{4\}$ | Z | | $P_{3,0} = 1$ |
| | 1 | | $\varnothing$ | $(0, 0)$ | $p_{3:\varnothing} = p_{3,1}$ |
| | 2 | | $\{4\}$ | $(0, 1)$ | $p_{3:\{4\}} = p_{3,2}$ |

## 4.2 The new node-based BAT

To integrate the strength of the BAT proposed in [] for the AMIN reliability problem, a novel node-based BAT by combining the novel state concepts discussion in Sections 4.1 and the BAT proposed in [] is proposed here. The general pseudo code for the proposed node-based BAT is presented below in calculating $R_{1:\{|V|\}}$:

**Node-based BAT**

**Input:** The MIN $G(V, E, \mathbf{D})$, $V_i$, and $S_{i,\text{Deg}(i)}$, for all $\text{Deg}(i)$ is the degree of node $i \in V$.

**Output:** The success probability $R_{1:\{|V|\}}$ that node 1 is connected to each node in $\tau$.

**STEP 0.** Let $X(1) = 1$, $X(i) = 0$ for all $i = 2, 3, \ldots, |V|$, $\text{Deg}(|V|) = v = 1$, and $R = 0$.

**STEP 1.** If $X$ is a feasible state vector, let $R = R + R(X)$.

**STEP 2.** If $X(v) < \text{Deg}(v)$, let $X(v) = X(v) + 1$, $v = 1$, and go to STEP 1.

**STEP 3.** Let $X(v) = 0$ if $v > 1$ or $X(v) = 2$ if $v = 1$, $v = v + 1$, and go to STEP 4.

**STEP 4.** If $v = |V|$, halt and $R = R_{1:\{|V|\}}$. Otherwise, go to STEP 2.

STEP 0 initializes the state vector $X$ which is a zero vector except that the value of first element is 1, the temporary reliability $R$, and starts the whole procedure from node 1. STEP 1 is to calculate the reliability of the current state vector if it is feasible. STEP 2 is to check whether the value (state) of current visiting node reaches the maximum. If not, increases the value of its state by one and returns to STEP 1 after resetting $v = 1$. Otherwise, go to STEP 4. STEP 4 is the proposed stopping criteria which is more efficient than the original BAT [].

There are $(2^{|\text{Deg}(i)|} + 1)$ states for each node $i \in V$ as mentioned in Section 4.1. It takes $O(|V|)$ to verify whether $X$ is a feasible state vector and calculate $R(X)$ in STEP 1. Hence, in the worst cases, the time complexity is $O(2^{\text{Deg}(1)} \cdot 2^{\text{Deg}(2)} \cdots \cdot 2^{\text{Deg}(|V|-1)}) \cdot O(|V|) = O(|V| \, 2^{|E|})$ which is exactly the same to that in UGFM discussed in Section 3.2 []. Also, the required space is only $O(|V|)$ to store $X$ and it is much smaller than that of UGFM which needs $O(2^{|E|})$.

Thus, from the above, the proposed new BAT is more economic in required space than that in the UGFM.

### 4.3 Demonstration Example

The run time growths exponentially with problem size in evaluating the AMIN reliability of which it is an NP-Hard problem. Therefore, the suggested methods cannot be applied for large-scale networks. To avoid this inherent computational difficulty, instead of giving nearly large AMINs, the small-size benchmark AMIN shown in Fig. 1, is thus applied to exemplify the proposed node-based BAT step-by-step to calculate AMIN reliability:

**STEP 0.** Let $X = (1, 0, 0, 0)$, $\text{Deg}(|V|) = \text{Deg}(4) = v = 1$, and $R = 0$.

**STEP 1.** Because $X = (1, 0, 0, 0)$ cannot send information from nodes 1 to 4, i.e., $X$ is not a feasible state vector, go to STEP 2.

**STEP 2.** Because $X(1) = 1 < \text{Deg}(1) = 2^{\text{Deg}(v_1)} = 2^2 = 4$, let $X(1) = X(1) + 1 = 2$ and go to STEP 1.

**STEP 1.** Because $X = (2, 0, 0, 0)$ cannot send information from nodes 1 to $|V|$, i.e., $X$ is not a feasible state vector, go to STEP 2.

**STEP 2.** Because $X(1) = 2 < 2^{\text{Deg}(v_1)} = 4$, let $X(1) = X(1) + 1 = 3$ and go to STEP 1.

**STEP 1.** Because $X = (3, 0, 0, 0)$ cannot send information from nodes 1 to $|V|$, i.e., $X$ is not a feasible state vector, go to STEP 2.

**STEP 2.** Because $X(1) = 3 < 2^{\text{Deg}(v_1)} = 4$, let $X(1) = X(1) + 1 = 4$ and go to STEP 1.

**STEP 1.** Because $X = (4, 0, 0, 0)$ cannot send information from nodes 1 to $|V|$, i.e., $X$ is not a feasible state vector, go to STEP 2.

**STEP 2.** Because $X(1) = 2^{\mathrm{Deg}(v_1)} = 4$, go to STEP 3.

**STEP 3.** Because $v = 1$, let $X(1) = 2$, $v = v + 1 = 2$, and go to STEP 4.

**STEP 4.** Because $v = 2 < |V| = 4$, go to STEP 2.

**STEP 2.** Because $X(2) = 0 < 2^{\mathrm{Deg}(v_2)} = 4$, let $X(2) = X(2) + 1 = 1$ and go to STEP 1.

**STEP 1.** Because $X = (2, 1, 0, 0)$ cannot send information from nodes 1 to $|V|$, i.e., $X$ is not a feasible state vector, go to STEP 2.

**STEP 2.** Because $X(1) = 2 < 2^{\mathrm{Deg}(v_1)} = 4$, let $X(1) = X(1) + 1 = 3$ and go to STEP 1.

**STEP 1.** Because $X = (3, 1, 0, 0)$ cannot send information from nodes 1 to $|V|$, i.e., $X$ is not a feasible state vector, go to STEP 2.

**STEP 2.** Because $X(1) = 3 < 2^{\mathrm{Deg}(v_1)} = 4$, let $X(1) = X(1) + 1 = 4$ and go to STEP 1.

**STEP 1.** Because $X = (4, 1, 0, 0)$ cannot send information from nodes 1 to $|V|$, i.e., $X$ is not a feasible state vector, go to STEP 2.

**STEP 2.** Because $X(1) = 2^{\mathrm{Deg}(v_1)} = 4$, go to STEP 3.

**STEP 3.** Because $v = 1$, let $X(1) = 2$, $v = v + 1 = 2$, and go to STEP 4.

**STEP 4.** Because $v = 2 < |V| = 4$, go to STEP 2.

$$\vdots$$
$$\vdots$$

**STEP 1.** Because $X = (2, 3, 0, 0)$ can send information from node 1 to $S_{1,2} = \{2\}$; from node 2 to $S_{2,3} = \{4\} = \{|V|\}$ without any redundant nodes, i.e., $X$ is a feasible state vector, let let $R = R + R(X) = R(X) = p_{1,2}p_{2,3}$ and go to STEP 2.

**STEP 2.** Because $X(1) = 2 < 2^{\mathrm{Deg}(v_1)} = 4$, let $X(1) = X(1) + 1 = 3$ and go to STEP 1.

**STEP 1.** Because $X = (3, 3, 0, 0)$ cannot send information from nodes 1 to $|V|$, i.e., $X$ is not a feasible state vector, go to STEP 2.

**STEP 2.** Because $X(1) = 3 < 2^{\mathrm{Deg}(v_1)} = 4$, let $X(1) = X(1) + 1 = 4$ and go to STEP 1.

**STEP 1.** Because $X = (4, 3, 0, 0)$ can send information from nodes 1 to $|V|$, i.e., $X$ is not a feasible state vector, go to STEP 2.

:

:

The procedure and results of the proposed BAT in calculating the one-to-many reliability on the MIN

shown in Fig. 1 are summarized in Table ?.

**Table ?.** The states and the corresponding codes of each node.

| code \ node | 1 | 2 | 3 | 4 |
|---|---|---|---|---|
| 0 | | | | |
| 1 | $\varnothing$ | $\varnothing$ | $\varnothing$ | |
| 2 | $\{2\}$ | $\{3\}$ | $\{4\}$ | |
| 3 | $\{3\}$ | $\{4\}$ | | |
| 4 | $\{2, 3\}$ | $\{3, 4\}$ | | |

**Table ?.** The feasible state vectors obtained from the proposed BAT.

| $i$ | $j$ | $X_j$ | $P(X_i)$ | $P(X_i)$ | Remark |
|---|---|---|---|---|---|
| 11 | 1 | $(2, 3, 0)$ | $p_{1,2}p_{2,3}$ | $p_{1:\{2\}}p_{2:\{4\}}$ | Fig. ?(a) |
| 28 | 2 | $(4, 3, 1)$ | $p_{1,4}p_{2,3}p_{3,1}$ | $p_{1:\{3,4\}}p_{2:\{4\}}p_{3,\varnothing}$ | Fig. ?(b) |
| 29 | 3 | $(2, 4, 1)$ | $p_{1,2}p_{2,4}p_{3,1}$ | $p_{1:\{2\}}p_{2:\{3,4\}}p_{3,\varnothing}$ | Fig. ?(c) |
| 31 | 4 | $(4, 4, 1)$ | $p_{1,4}p_{2,4}p_{3,1}$ | $p_{1:\{3,4\}}p_{2:\{3,4\}}p_{3,\varnothing}$ | Fig. ?(d) |
| 33 | 5 | $(3, 0, 2)$ | $p_{1,3}p_{2,0}p_{3,2}$ | $p_{1:\{3\}}p_{3:\{4\}}$ | Fig. ?(e) |
| 37 | 6 | $(4, 1\ 2)$ | $p_{1,4}p_{2,1}p_{3,2}$ | $p_{1:\{3,4\}}p_{2,1}p_{3:\{4\}}$ | Fig. ?(f) |
| 38 | 7 | $(2, 2, 2)$ | $p_{1,2}p_{2,2}p_{3,2}$ | $p_{1:\{2\}}p_{2:\{3\}}p_{3:\{4\}}$ | Fig. ?(g) |
| 40 | 8 | $(4, 2, 2)$ | $p_{1,4}p_{2,2}p_{3,2}$ | $p_{1:\{3,4\}}p_{2:\{3\}}p_{3:\{4\}}$ | Fig. ?(h) |
| 43 | 9 | $(4, 3, 2)$ | $p_{1,4}p_{2,3}p_{3,2}$ | $p_{1:\{3,4\}}p_{2:\{4\}}p_{3:\{4\}}$ | Fig. ?(i) |
| 44 | 10 | $(2, 4, 2)$ | $p_{1,2}p_{2,4}p_{3,2}$ | $p_{1:\{2\}}p_{2:\{3,4\}}p_{3:\{4\}}$ | Fig. ?(j) |
| 46 | 11 | $(4, 4, 2)$ | $p_{1,4}p_{2,4}p_{3,2}$ | $p_{1:\{3,4\}}p_{2:\{3,4\}}p_{3:\{4\}}$ | Fig. ?(k) |

There are 76 state vectors found in the BAT procedure and 11 among them are feasible state vectors.

If the state probability of each node is equal, i.e., $p_{i,k} = 1/[2^{\text{Deg}(i)}]$, the total probability is 0.468750.

## 5. EXPERIMENTAL RESULTS

To demonstrate the benefit of the proposed new BAT, both BAT and UGFM are compared and test here. Note that only UGFM is implemented to compare with BAT is because of that UGFM is the main algorithm for MIN the reliability problems.

To ensure a fair comparison, both the proposed BAT and UGFM [38] are implemented under the same computational environment: designed by means of Windows 10 64-bit DEV C++, applied on an Intel Core i7-5960X CPU @ 3.00GHz 3.00GHz with 16 GB memory, and measured the computer execution time in CPU seconds. Moreover, the maximal space of the UGFM to save all feasible sub-vectors is 1,900,000 in a double-decision flow format. Note that there is no need to have such larger memory for the BAT.

Without loss of generality, the state reliability of each node is equal to $1/2^{\text{DEG}(v)}$ for all $v \in V$, and node 1 and $|V|$ are the source node and sink node, respectively. Note that the number of states is $2^{\text{DEG}(v)}$ for each node $v$.

The MIN is mainly applied to information networks, e.g., wireless local area networks (WLANs) for various or many clients. These networks are very dense in common, thus, we proposed semi-complete MINs of which the degree of each nodes is maximum such that node $i$ is connected to all nodes $j$ where $i < j$ to test the performance of both BAT and UGFM.

The number of nodes, i.e., $|V|$, of these semi-complete MINs are started from $|V| = 5$ and increased by one for next MIN until one of BAT and UGFM failed to calculate its reliability. All obtained results are listed in Table ?, where $N_{\text{all}}$, N, $N_\bullet$, and $T_\bullet$ are the number of all possible state vectors, the number of all feasible state vectors, the number of state vectors found and tested by algorithm ●, and the total runtime by algorithm ●, respectively.

In Table 5, the bold numbers represent the best results between the proposed BAT and UGFM. Both the proposed BAT and UGFM obtained the same number of feasible state vectors, i.e., N, for each test. The test is stopped at the AMIN with $|V| = 8$ because the UGFM is failed already. Hence, UGFM can only solves these problems with $|V| < 8$; on the contrary, the proposed BAT still can solve these problems with more than 8 nodes.

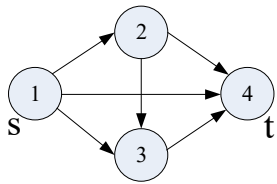

In UGFM, all state sub-vectors must be kept in computer memory until these sub-vectors became full vectors and this causes the system crashes if the number of sub-vectors overflows. Note that extra computer memory, e.g., internet, hard disk space, etc., to stock sub-vectors at the cost of longer runtime spent in moving, copying, and saving. However, memory requires exponentially to overcome this obstacle which is impossible even for middle-scale problems for the UGFM.

**Table 4.** Comparison of BAT and UGFM [16].

| $|V|$ | $|E|$ | $N_{all}$ | N | $R_{1:\{|V|\}}$ | $T_{BAT}$ | $T_{UGFM}$ | $N_{BAT}$ | $N_{UGFM}$ |
|---|---|---|---|---|---|---|---|---|
| 5 | 10 | 2,295 | 388 | 0.821289 | 0.000000 | 0.000000 | 2,025 | **705** |
| 6 | 15 | 75,735 | 11,164 | 0.884979 | **0.000000** | 0.002000 | 71,145 | **19,145** |
| 7 | 21 | 4,922,775 | 667,396 | 0.928662 | 0.279000 | **0.102000** | 4,771,305 | **1,117,441** |
| 8 | 28 | 635,037,975 | 81,974,044 | 0.957076 | **43.655998** | N/A | **625,192,425** | N/A |
| 9 | 36 | 163,204,759,575 | 20,483,270,788 | 0.974799 | **13,890.328125** | N/A | **161,934,683,625** | N/A |

Because $N_{BAT} > N_{UGFM}$ for $|V|$ = 5, 6, 7, it seems that UGFM is more efficient than the proposed BAT for smaller-scale problems. However, the difference is not that important from the runtime because both are less than 1 second for $|V|$ = 5, 6, 7. Also, the runtime is increased exponentially which is met the NP-Hard characteristic of the MIN reliability problems.

From Table ?, the larger $V$, the greater values of N and $R_{1:\{|V|\}}$. Hence, the strategy to have a reliable high-dense AMIN is to increase its size and this phenomenon also can be confirmed from the real-life applications, e.g., WLAN and 6G/4G, etc. An interesting observation is that N = $N_{BAT}$ = 16,777,216 for $|V|$ = 8, thus, each state vector is feasible for $|V| \geq 8$. Hence, it is impossible to fathom any semi-state vectors in the UGFM or to reduce any state vector in all related algorithms. The above observation results in a very important concept and is useful to develop any new algorithm or improve any existing algorithms for the high-dense AMIN reliability problems in the future. It also helpful to increase more connections and more customers to have a more reliable networks in running high-dense network related business, e.g., social networks, 4G/5G, WLAN, etc.

Another tricky part is that the numbers of N for both $|V|$ = 8 and 9 is equal. The reason is that the data type is a single-precision float-point format to count the number of all possible state vectors, i.e., N. The number of N will get an overflow answer if it is larger a certain number because the predefined limitation from the C/C++. Note that these results of $R_{1:\{|V|\}}$ and the runtime are without the overflow problems.

From here we can also knowledge that not only the computer memory is a limitation but also the computer language. Hence, the traditional UGFM needs to have all state vectors is an impossible mission. Contrary, the proposed BAT only requires an $|V|$-tupe array to store the current state vector without needing to have exponential computer memory to save all vectors. Thus, it further confirms that the advantage of the proposed BAT in the view from the computer memory.

## 6. CONCLUSIONS

Network reliability has been implement as an important maeasure index extensively in many practical applications. UGFM plays a noteworthy role in calculating the exact AMFN reliability. However, this study reveals a computer memory overflow problem in UGFM even for a small size dense AMIN as in Section 5. In this study, our goal is to develop a new technique for generalizing the existing BAT to overcome such serious memory overflow problem in calculating the exact AMIN reliability.

From the time complexity provided in Section 4.2, the proposed node-based BAT algorithm is as efficient as that of the UGFM. Also, in Section 4.2, from the pseudo code, the proposed BAT is easy to program and the major computer memory is only to store the current state vector, i.e., efficient in managing computer memory. From the same example provided for both UGFM (in Section 3.2) and the proposed BAT (in Section 4.3), the proposed BAT is simpler to understand and less tedious than that of UGFM.

More important, the computer memory requires for the proposed BAT is only $O(|V|)$ which is much less than that of UGFM for solving the AMIN reliability problems. From the experiments on high-dense AMIN in Section 5, the proposed BAT further confirms that it outperforms the UGFM without having memory overflow problem. Moreover, in out experiments, we find some import properties which are helpful in developing algorithms or running related businesses:

1. The larger size, the more reliable AMIN. Without reliable connection in commutations, no matter the price is less expensive, customers will move to another company. Also, the simple way to have more reliable commutations is to have more users and this may be different to our common sense.

2. All state vectors are feasible. Hence, no need to have simplifications to reduce the number of feasible state vectors and the computer memory control is more critical and important than the efficiency of algorithms for larger-size AMIN reliability problems.

Thus, the proposed node-based BAT is more attractive than the UGFM for high dense AMIN in simplicity, efficiency, and computer memory. The reliability of AMIN seems too difficult to calculate for larger-size AMINs. However, the exact reliability is still the basis to evaluate the performance of AMINs correctly and even these solutions are easier to obtain from simulations. There is a need to combine BAT and simulation to have advantages as much as possible and the shortcomings as less as possible of both BAT and simulation in the future works.

**ACKNOWLEDGEMENTS**

This research was supported in part by the Ministry of Science and Technology of Taiwan, under grant MOST 107-2221-E-007-072-MY3.

### 4.2 Flexible State Vector and Probability

State vector records the sequence of the information spread from node 1 to target nodes. A feasible state vector is a vector such that the information is able to be spread from the source node, i.e., node 1, to target nodes. For example, the state vector that corresponding to the information spread from nodes 1 to 2 to 3, and stop in Fig. 1 is not a feasible state vector.

In the proposed new BAT, the number of coordinates in a state vector is not a constant and it depends on how the information is spread. Moreover, the position of each node in the vector is also not fixed. Such vectors are called flexible state vectors here.

To recognize which node is with which state in the flexible state vector, each coordinate is denoted by a two-tuple sub-vector such that the first tuple is the current node and the second tuple is the node state label of the current shown in the first tuple, i.e., the notation $(1, 4)$ denotes that node 1 is in state (label) 4, i.e., $S_{1,4} = \{2, 3\}$.

In the flexible state vector, the first tuple in the first two-tuple sub-vector is always the source node, i.e., node 1, and following the second tuple is the node state label of node 1. The first tuple in the second two-tuple sub-vector is always the node in the second tuple of the first two-tuple sub-vector that received the information from node 1. In the same way, the first tuple in the $k$th two-tuple sub-vector must received the information from at least one node appeared in the second tuple of the two-tuple sub-vectors that in precedence of the $k$th two-tuple sub-vector.

For example, $X_2 = ((1, 4), (2, 1), (3, 2))$ is a state vector including nodes 1, 2, and 3, with node state labels 4, 1, and 2, respectively. The state vector $X_2$ represents the information is flew from node 1 to its node state label 4, i.e., $S_{1,4} = \{2, 3\}$; then from node 2 to $S_{2,1} = \varnothing$, and from node 3, to $S_{3,2} = \{4\}$. The corresponding probabilities from node 1 to $S_{1,4} = \{2, 3\}$, from node 2 to $S_{2,1} = \varnothing$, from node 3, to $S_{3,2} = \{4\}$ are $p_{1:\{2,3\}}$, $p_{2,\varnothing}$, and $p_{3:\{4\}}$, respectively. All the above spread must be occurrent to have $X_2 = ((1, 4), (2, 1), (3, 2))$. Hence, $\Pr(X_2) = p_{1:\{2,3\}} p_{2,\varnothing} p_{3:\{4\}}$.

Analogy, all feasible state vectors and their probabilities are shown in Table ?.

Table ?. The state vectors and their probabilities.

| $i$ | $X_i$ | $\Pr(X_i)$ |
|---|---|---|
| 1 | $((1, 1), (2, 2), (3, 2))$ | $p_{1,1}p_{2,2}p_{3,2} = p_{1:\{2\}}p_{2:\{3\}}p_{3:\{4\}}$ |
| 2 | $((1, 2), (2, 3))$ | $p_{1,2}p_{2,3,} = p_{1:\{2\}}p_{2:\{5\}}$ |
| 3 | $((1, 2), (2, 4), (3, 1))$ | $p_{1,2}p_{2,4}p_{3,1} = p_{1:\{2\}}p_{2:\{3,5\}}p_{3,\varnothing}$ |
| 4 | $((1, 2), (2, 4), (3, 2))$ | $p_{1,2}p_{2,4}p_{3,2} = p_{1:\{2\}}p_{2:\{3,5\}}p_{3:\{4\}}$ |
| 5 | $((1, 3), (3, 2))$ | $p_{1,3}p_{3,2} = p_{1:\{3\}}p_{3:\{4\}}$ |
| 6 | $((1, 4), (2, 1), (3, 2))$ | $p_{1,4}p_{2,1}p_{3,2} = p_{1:\{2,3\}}p_{2,\varnothing}p_{3:\{4\}}$ |
| 7 | $((1, 4), (2, 2), (3, 2))$ | $p_{1,4}p_{2,2}p_{3,2} = p_{1:\{2,3\}}p_{2:\{3\}}p_{3:\{4\}}$ |
| 8 | $((1, 4), (2, 3), (3, 1))$ | $p_{1,4}p_{2,3}p_{3,1} = p_{1:\{2,3\}}p_{2:\{5\}}p_{3,\varnothing}$ |
| 9 | $((1, 4), (2, 3), (3, 2))$ | $p_{1,4}p_{2,3}p_{3,2} = p_{1:\{2,3\}}p_{2:\{5\}}p_{3:\{4\}}$ |
| 10 | $((1, 4), (2, 4), (3, 1))$ | $p_{1,4}p_{2,4}p_{3,1} = p_{1:\{2,3\}}p_{2:\{3,5\}}p_{3,\varnothing}$ |
| 11 | $((1, 4), (2, 4), (3, 2))$ | $p_{1,4}p_{2,4}p_{3,2} = p_{1:\{2,3\}}p_{2:\{3,5\}}p_{3:\{4\}}$ |

The corresponding diagram of each feasible state vector is depictured in Fig. ? to show how the spread of flow.

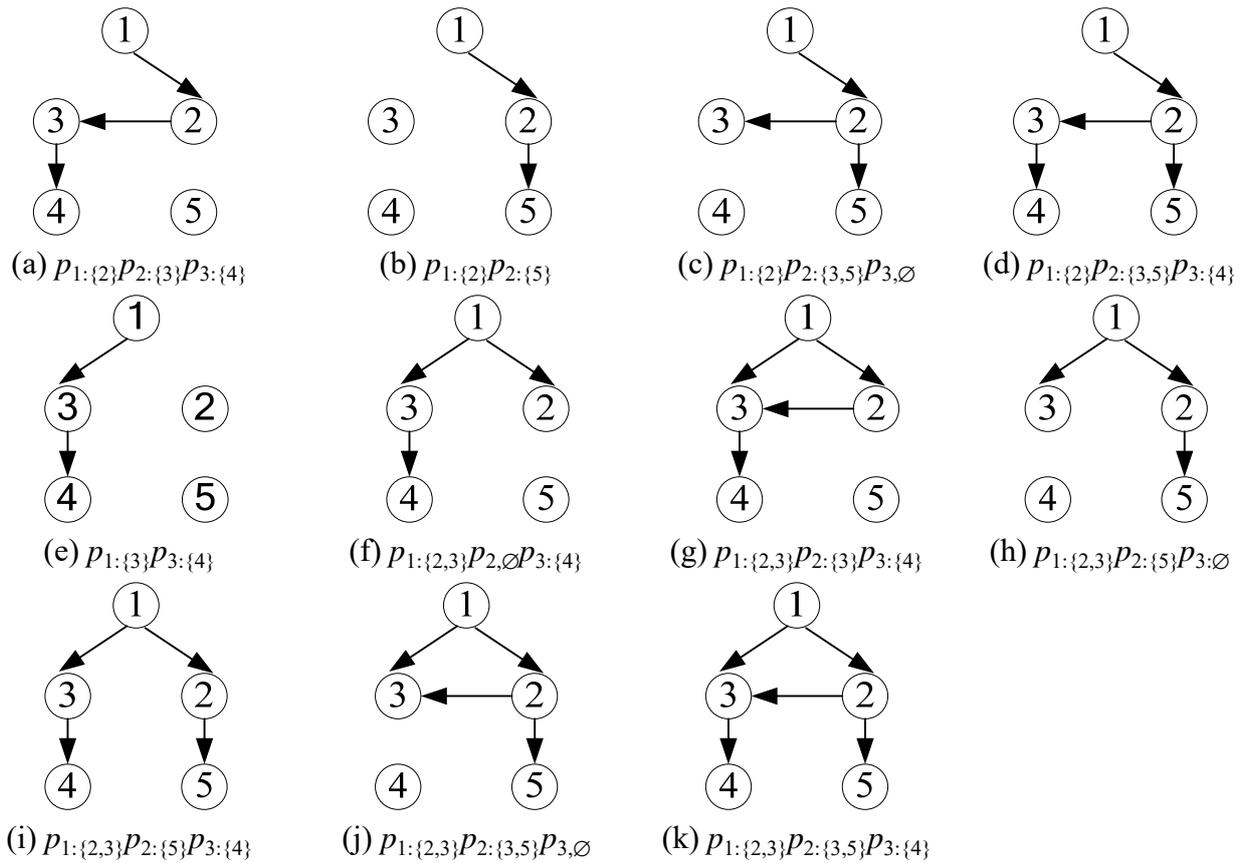

(a) $p_{1:\{2\}}p_{2:\{3\}}p_{3:\{4\}}$

(b) $p_{1:\{2\}}p_{2:\{5\}}$

(c) $p_{1:\{2\}}p_{2:\{3,5\}}p_{3,\varnothing}$

(d) $p_{1:\{2\}}p_{2:\{3,5\}}p_{3:\{4\}}$

(e) $p_{1:\{3\}}p_{3:\{4\}}$

(f) $p_{1:\{2,3\}}p_{2,\varnothing}p_{3:\{4\}}$

(g) $p_{1:\{2,3\}}p_{2:\{3\}}p_{3:\{4\}}$

(h) $p_{1:\{2,3\}}p_{2:\{5\}}p_{3,\varnothing}$

(i) $p_{1:\{2,3\}}p_{2:\{5\}}p_{3:\{4\}}$

(j) $p_{1:\{2,3\}}p_{2:\{3,5\}}p_{3,\varnothing}$

(k) $p_{1:\{2,3\}}p_{2:\{3,5\}}p_{3:\{4\}}$

**Figure 2.** Subgraphs corresponding to feasible state vectors in Table ?.

From the above, an important advantage of the proposed flexible state vector is that how the information spread from node 1 to target nodes can be easily identified in the state vector.

## DEL

s and subnet-UGF which is based on each state of nodes and sub-networks is correct [31-33, 38, 42-44]. The number of states in $u(i)$ is $O(2^{\text{outdegree}(i)})$, where outdegree($i$) is the number of arcs in $\{e_{ij}|$ for all $e_{ij} \in E\}$ [33, 38, 44]. in AMIN [31-33]. The UGFM

The UGFM is generalized for multistate network reliability by Yeh [38]. Moreover, Yeh extended the existing UGFM as the convolution UGFM [33] to solve one-to-all-target-subset reliability evaluations of the AMIN problem with improved time complexity from $O(2^{2|E| - |T|})$ [38, 44] to $O(|V| \prod_{i=1}^{t} [\text{outdegree}(i) + 2])$ [33], where $T = \{t+1, t+2, ..., /V/\}$ is the set of sink nodes, which are the target set that is unable to transmit a signal and $t$ is the cardinality of the set of non-target (non-sink) nodes, i.e. $t=|V-T|$ [33, 38, 44]. However,

An application of the operator to acyclic consecutively connected networks with multi-state elements was first proposed by Levitin [17], and improved by Yeh [20], [21].

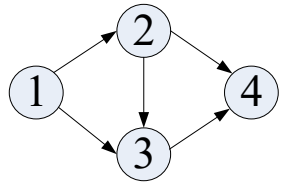

Fig. 1. An AMIN example.

**STEP 2.** Because $X(2) = 1 < 2^{\text{Deg}(v_2)} = 4$, let $X(2) = X(2) + 1 = 2$ and go to STEP 1.

**STEP 2.** Because $X(1) = 2 < 2^{\text{Deg}(v_1)} = 4$, let $X(1) = X(1) + 1 = 3$ and go to STEP 1.

**STEP 1.** Because $X = (3, 2, 0, 0)$ cannot send information from nodes 1 to $|V|$, go to STEP 2.

**STEP 1.** Because $X = (3, 2, 0, 0)$ can send information from nodes 1 to $|V|$, i.e., from nodes 1 to $S_{1,3} = \{3\}$ $X$ is not a feasible state vector, go to STEP 2.

**STEP 1.** Because $X = (4, 0, 0, 0)$ cannot send information from nodes 1 to $|V|$, i.e., $X$ is not a feasible state vector, go to STEP 2.

**STEP 2.** Because $X(1) = 2^{\text{Deg}(v_1)} = 4$, go to STEP 3.

**STEP 4.** Because $v = 1 \neq |V| = 4$, go to STEP 5.

**STEP 5.** Let and go to STEP 1.

**STEP 5.** Let $v = 1$ and go to STEP 1.

**STEP 1.** Because $X(1) = 2 < \text{Deg}(1) = 4$, let $X(1) = X(1) + 1 = 3$, and go to STEP 3.

**STEP 3.** Because $X(|V|) = X(4) = 0$, go to STEP 4.

**STEP 4.** Because $X = (3, 0, 0, 0)$ cannot send information from nodes 1 to $|V|$, i.e., $X$ is not a feasible state vector, go to STEP 5.

**STEP 5.** Let $v = 1$ and go to STEP 1.

**⋮**

:

**STEP 3.** Let $c = c + 1 = 2$, $v_c = 2$ be the only node in $L_1 = \{2\}$, $s_c = 2$ because $|L_1| = 1$ and $\tau_1 = \varnothing$, and go to STEP 1. Note that $X = ((1, 2), (2, 2))$ now.

**STEP 1.** Because $L_2 = [L_1 - \{v_2\}] \cup S_{2,2} = S_{2,2} = \{3\} \neq \varnothing$, go to STEP 2.

**STEP 2.** Because $L_2 \cap T = \varnothing$, let $\tau_2 = \tau_1 = \varnothing$ and go to STEP 3.

**STEP 3.** Let $c = c + 1 = 3$, $v_{3,} = 3$, be the only node in $L_{c-1} = L_2 = \{3\}$, $s_c = 2$ because $|L_2| = 1$ and $\tau_2 = \varnothing$, and go to STEP 1. Note that $X = ((1, 2), (2, 2), (3, 2))$ now.

**STEP 1.** Because $L_{3,} = [L_2 - \{v_3\}] \cup S_{3,2} = S_{3,2} = \{4\} \neq \varnothing$, go to STEP 2.

**STEP 2.** Because $L_{3,} \cap T = \{4\} = L_3$, $X$ is a feasible state vector (see Fig. 2a) and go to STEP 6.

**STEP 6.** Let $\Pr(X) = p_{1,2}p_{2,2}p_{3,2} = p_{1:\{2\}}p_{2:\{3\}}p_{3:\{4\}}$, $R(\tau_2 \cup \{4\}) = R(\{4\}) + \Pr(X) = p_{1:\{2\}}p_{2:\{3\}}p_{3:\{4\}}$, and go to STEP 4.

**STEP 5.** Because $c = 3, > 1$, let $c = c - 1 = 2$ and go to STEP 4.

**STEP 4.** Because $s_c = 2 < 2^{\text{Deg}(v_2)} = 2^2$, let $s_c = s_c + 1 = 3$, and go to STEP 1. Note that $X = ((1, 2), (2, 3))$ now.

**STEP 1.** Because $L_2 = [L_1 - \{v_2\}] \cup S_{2,3} = S_{2,3} = \{5\} \neq \varnothing$, go to STEP 2.

**STEP 2.** Because $L_2 \cap T = \{5\} = L_2$, $X$ is a feasible state vector (see Fig. 2b) and go to STEP 6.

**STEP 6.** Let $\Pr(X) = p_{1,2}p_{2,3,} = p_{1:\{2\}}p_{2:\{5\}}$, $R(\tau_1 \cup \{5\}) = R(\{5\}) + \Pr(X) = p_{1:\{2\}}p_{2:\{5\}}$, and go to STEP 4.

**STEP 4.** Because $s_c = 3, < 2^{\text{Deg}(v_2)} = 2^2$, let $s_c = s_c + 1 = 4$ and go to STEP 1. Note that $X = ((1, 2), (2, 4))$ now.

**STEP 1.** Because $L_2 = [L_1 - \{v_2\}] \cup S_{2,4} = S_{2,4} = \{3, 5\} \neq \varnothing$, go to STEP 2.

**STEP 2.** Because $L_2 \cap T = \{5\}$, let $L_2 = L_2 - \{5\} = \{3\}$, $\tau_2 = \tau_1 \cup \{5\} = \{5\}$ and go to STEP 3.

**STEP 3.** Let $c = c + 1 = 3$, $v_3 = 3$, be the only node in $L_{c-1} = L_2 = \{3\}$, $s_c = 1$, $X$ is a feasible state vector (see Fig. 2c), and go to STEP 6 because $|L_2| = 1$ and $\tau_2 \neq \varnothing$. Note that $X = ((1, 2), (2, 4), (3, 1))$ now.

**STEP 6.** Let $\Pr(X) = p_{1,2}p_{2,4}p_{3,1} = p_{1:\{2\}}p_{2:\{3,5\}}p_{3:\varnothing}$, $R(\tau_2) = R(\{5\}) + \Pr(X) = p_{1:\{2\}}p_{2:\{5\}} + p_{1:\{2\}}p_{2:\{3,5\}}p_{3:\varnothing}$, and go to STEP 4.

**STEP 4.** Because $s_c = 1 < 2^{\text{Deg}(v_3)} = 2$, let $s_c = s_c + 1 = 2$ and go to STEP 1. Note that $X = ((1, 2), (2, 4), (3, 2))$ now.

**STEP 1.** Because $L_{3,} = [L_2 - \{v_3\}] \cup S_{3,2} = S_{3,2} = \{4\} \neq \varnothing$, go to STEP 2.

**STEP 2.** Because $L_{3,} \cap T = \{4\} = L_3$, $X$ is a feasible state vector (see Fig. 2d) and go to STEP 6.

**STEP 6.** Let $\Pr(X) = p_{1,2}p_{2,4}p_{3,2} = p_{1:\{2\}}p_{2:\{3,5\}}p_{3:\{4\}}$, $R(\tau_2 \cup \{4\}) = R(\{4, 5\}) + \Pr(X) = p_{1:\{2\}}p_{2:\{3,5\}}p_{3:\{4\}}$, and go to STEP 4.

$$\vdots$$

$$\vdots$$

The procedure and results of the proposed BAT in calculating the one-to-many reliability on the MIN shown in Fig. 1 are summarized in Table ?.

**Table ?.** Final results obtained from the proposed new BAT tested on Fig. 1[*].

| $c$ | $v_c$ | $s_c$ | $X$ | $S_{i,j}$ | $L_c$ | $L^*$ | $L_c^{\#}$ | $\tau_c$ | | Remark |
|---|---|---|---|---|---|---|---|---|---|---|
| 0 | | | | | $\{1\}$ | | | $\varnothing$ | | |
| 1 | 1 | 2 | $((1, 2))$ | $\{2\}$ | $\{2\}$ | | | $\varnothing$ | | |
| 2 | 2 | 2 | $((1, 2), (2, 2))$ | $\{3\}$ | $\{3\}$ | | | $\varnothing$ | | |
| 3 | 3 | 2 | $((1, 2), (2, 2), (3, 2))$ | $\{4\}$ | $\{4\}$ | | | | $p_{1:\{2\}}p_{2:\{3\}}p_{3:\{4\}}$ | Fig. 2a |
| 2 | 2 | 3 | ((1, 2), (2, 3)) | {5} | {5} | | | | $p_{1:\{2\}}p_{2:\{5\}}$ | Fig. 2b |
| 2 | 2 | 4 | ((1, 2), (2, 4)) | {3, 5} | {3, 5} | {5} | {3} | {5} | | |
| 3 | 3 | 1 | ((1, 2), (2, 4), (3, 1)) | $\varnothing$ | $\varnothing$ | | | {5} | $p_{1:\{2\}}p_{2:\{3,5\}}p_{3:\varnothing}$ | Fig. 2c |
| 3 | 3 | 2 | ((1, 2), (2, 4), (3, 2)) | {4} | {4} | | | {4, 5} | $p_{1:\{2\}}p_{2:\{3,5\}}p_{3:\{4\}}$ | Fig. 2d |
| 1 | 1 | 3 | ((1, 3)) | {3} | {3} | | | $\varnothing$ | | |
| 2 | 3 | 2 | ((1, 3), (3, 2)) | {4} | {4} | | | {4} | $p_{1:\{3\}}p_{3:\{4\}}$ | Fig. 2e |
| 1 | 1 | 4 | ((1, 4)) | {2, 3} | {2, 3} | | | $\varnothing$ | | |
| 2 | 2 | 1 | ((1, 4), (2, 1)) | $\varnothing$ | {3} | | | $\varnothing$ | | |
| 3 | 3 | 2 | ((1, 4), (2, 1), (3, 2)) | {4} | {4} | | | {4} | $p_{1:\{2,3\}}p_{2:\varnothing}p_{3:\{4\}}$ | Fig. 2f |
| 2 | 2 | 2 | ((1, 4), (2, 2)) | {3} | {3} | | | $\varnothing$ | | |
| 3 | 3 | 2 | ((1, 4), (2, 2), (3, 2)) | {4} | {4} | | | {4} | $p_{1:\{2,3\}}p_{2:\{3\}}p_{3:\{4\}}$ | Fig. 2g |
| 2 | 2 | 3 | ((1, 4), (2, 3)) | {5} | {3} | | | {5} | | |
| 3 | 3 | 1 | ((1, 4), (2, 3), (3, 1)) | $\varnothing$ | $\varnothing$ | | | {5} | $p_{1:\{2,3\}}p_{2:\{5\}}p_{3:\varnothing}$ | Fig. 2g |
| 3 | 3 | 2 | ((1, 4), (2, 3), (3, 2)) | {4} | {4} | | | {4, 5} | $p_{1:\{2,3\}}p_{2:\{5\}}p_{3:\{4\}}$ | Fig. 2i |
| 2 | 2 | 4 | ((1, 4), (2, 4)) | {3, 5} | {3, 5} | {5} | {3} | {5} | | |
| 3 | 3 | 1 | ((1, 4), (2, 4), (3, 1)) | $\varnothing$ | $\varnothing$ | | | {5} | $p_{1:\{2,3\}}p_{2:\{3,5\}}p_{3:\varnothing}$ | Fig. 2j |
| 3 | 3 | 2 | ((1, 4), (2, 4), (3, 2)) | {4} | {4} | | | {4, 5} | $p_{1:\{2,3\}}p_{2:\{3,5\}}p_{3:\{4\}}$ | Fig. 2k |

\*   $i = v_c$ and $j = s_c$ in $S_{i,j}$.

\#   $L_c = L_c - L^*$ based on STEP 2 if necessary.

The RSDP presented in [38] is currently the most popular and efficient SDP for MFN. RSDP was therefore selected for comparison with the proposed QIE. The best-known IET, called the Greedy-B&B-IE, was also proposed by the authors in [31]. The Greedy-B&B-IE is implemented based on the Branch-and-Bound technique (B&B) with special reduction technologies to reduce the number of intersections and decrease the multiplication for calculating the probability value [31]. The proposed QIE can be considered a simplified Greedy-B&B-IE [31] which replaces the complicated B&B with simple DFS. However, these special reduction technologies make Greedy-B&B-IE even more complicated. [31] If IET is to become widely implemented, it must be easily coded and easy to use, and to ensure a fair comparison, this study only compares QIE with RSDP, which does not include any reduction technology. However, these reduction technologies can be used in the proposed QIE.

**STEP 4.** Because $s_{3,} = 2^{\mathrm{Deg}(v_3)} = 2$, go to STEP 5.

**STEP 5.** Because $c = 3, > 1$, let $c = c - 1 = 2$ and go to STEP 4.

**STEP 4.** Because $s_2 = 2^{\mathrm{Deg}(v_2)} = 4$, go to STEP 5.

**STEP 5.** Because $c = 2 > 1$, let $c = c - 1 = 1$ and go to STEP 4.

**STEP 4.** Because $s_1 = 2 < 2^{\mathrm{Deg}(v_1)} = 4$, let $s_c = s_c + 1 = 3$, and go to STEP 1. Note that $X = ((1, 3))$ now.

**STEP 2.** Because $L_2 = \{3, 5\} \not\subset \tau_1 = \{4, 5\}$, $X$ is a feasible state vector and $\Pr(X) = p_{1,2}p_{2,3,} = p_{1:\{2\}}p_{2:\{5\}}$, $R(L_c) = R(\{5\}) + \Pr(X) = p_{1:\{2\}}p_{2:\{5\}}$, and go to STEP 4.

**STEP 5.** Because $c = 2 > 1$, let $c = c - 1 = 1$ and go to STEP 4.

**STEP 4.** Because $s_c = 2 < 2^{\mathrm{Deg}(v_1)} = 4$, let $s_c = s_c + 1 = 3$, and go to STEP 1. Note that $X = ((1, 3))$ now.

**STEP 1.** Because $L_1 = [L_0 - \{v_1\}] \cup S_{1,3,} = S_{1,3,} = \{3\} \neq \varnothing$, go to STEP 2.

**STEP 2.** Because $L_1 \not\subset \tau_0$, let $\tau_1 = \tau_0 - L_1 = \{4, 5\}$ and go to STEP 3.

**STEP 3.** Let $c = c + 1 = 2$, $v_2 = 3$, be the node with the smallest label in $L_{c-1} = L_1 = \{3\}$, $s_c = 1$, and go to STEP 1. Note that $X = ((1, 3), (3, 1))$ now.

**STEP 1.** Because $L_2 = [L_1 - \{v_2\}] \cup S_{3,1} = S_{3,1} = \varnothing$, go to STEP 4.

**STEP 4.** Because $s_c = 1 < 2^{\mathrm{Deg}(v_3)} = 2^1 = 2$, let $s_c = s_c + 1 = 2$ and go to STEP 1. Note that $X = ((1, 3), (3, 2))$ now.

**STEP 1.** Because $L_2 = [L_1 - \{v_3\}] \cup S_{3,2} = S_{3,2} = \{4\} \neq \varnothing$, go to STEP 2.

**STEP 2.** Because $L_2 = \{4\} \subseteq \tau_1 = \{4, 5\}$, $X$ is a feasible state vector and $\Pr(X) = p_{1,3}p_{3,2} = p_{1:\{3\}}p_{3:\{4\}}$, $R(L_c) = R(\{4\}) + \Pr(X) = p_{1:\{2\}}p_{2:\{5\}} + p_{1:\{3\}}p_{3:\{4\}}$, and go to STEP 4.

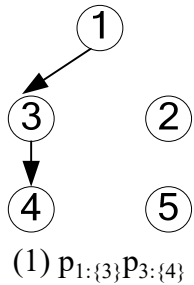

(1) $p_{1:\{3\}}p_{3:\{4\}}$

**STEP 4.** Because $s_c = 2^{\mathrm{Deg}(v_2)} = 2$, go to STEP 5.

**STEP 5.** Because $c = 2 > 1$, let $c = c - 1 = 1$ and go to STEP 4.

**STEP 4.** Because $s_c = 3, < 2^{\mathrm{Deg}(v_1)} = 4$, let $s_c = s_c + 1 = 4$ and go to STEP 1. Note that $X = ((1, 4))$ now.

**STEP 1.** Because $L_c = [L_0 - \{v_1\}] \cup S_{1,4} = S_{1,4} = \{2, 3\} \neq \varnothing$, go to STEP 2.

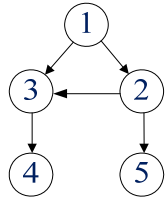

**STEP 0.** Let $L_0 = \{1\}$, the coordinate index $c = 1$, the node label $v_c = 1$, the node state label $s_c = 1$, $L_c = S_{1,1}$, and the current reliability $R = 0$.

**STEP 1.** Let $L_c = [L_{c-1} - \{v_c\}] \cup S_{i,j}$, where $i = v_c$ and $j = s_c$. If $L_c = \varnothing$, go to STEP 4.

**STEP 2.** If $\tau_c \subseteq L_c$, $X = ((v_1, s_1), (v_2, s_2), \ldots, (v_c, s_c))$ is a feasible state vector, let $R = R + \Pr(X)$ and go to STEP 4, where $\Pr(X) = \displaystyle\prod_{k=1}^{c} p_{v_k, s_k}$ . Otherwise, let $\tau_c = \tau_{c-1} - L_c$ and go to STEP 3

**STEP 3.** Let $c = c + 1$ and $v_c$ be the node with the smallest label in $L_{c-1}$.

**STEP 4.** If $s_c < 2^{\mathrm{Deg}(v_c)}$, let $s_c = s_c + 1$ and go to STEP 5. Otherwise, go to STEP 5.

**STEP 5.** If $c > 1$, let $c = c - 1$ and go to STEP 4. Otherwise, halt and $R_{1:\tau} = R$.

~~**STEP 4.** Let $c = c + 1 = 3$, $v_3 = 3$, be the node with the smallest label in $L_{c-1} = L_2 = \{3\}$, and go to STEP 1. Note that $X = ((1, 2), (2, 2), (3, 1))$ now.~~

~~**STEP 5.** Because $s_c = 2^{\mathrm{Deg}(v_c)} = 2^1 = 2$, go to STEP 6. Note that $X = ((1, 2), (2, 2), (3, 2))$ now.~~

~~**STEP 6.** Because $c = 3, > 1$, let $c = c - 1 = 2$ and go to STEP 5.~~

~~**STEP 5.** Because $s_c = 2 < 2^{\mathrm{Deg}(v_c)} = 2^2 = 4$, let $s_c = s_c + 1 = 3$, and go to STEP 1. Note that $X = ((1, 2), (2, 3))$ now.~~

~~**STEP 4.** Because $s_c = 1 < 2^{\mathrm{Deg}(v_c)} = 2^2 = 4$, let $s_c = s_c + 1 = 2$, and go to STEP 5.~~

~~**STEP 3.** Let $c = c + 1$, $v_c$ be the node with the smallest label in $L_{c-1}$ and go to STEP 5.~~

~~**STEP 4.** If $s_c < 2^{\mathrm{Deg}(v_c)}$, let $s_c = s_c + 1$ and go to STEP 5. Otherwise, go to STEP 6.~~

**STEP 5.** Let $L_e = [L_{e-1} - \{v_e\}] \cup S_{i,j_i}$, where $i = v_e$ and $j = s_e$, $\tau_e = \tau_{e-1} - \{v_e\}$, and go to STEP 1.

**STEP 6.** If $e > 1$, let $e = e - 1$ and go to STEP 4. Otherwise, halt and $R_{1-e} = R$.

In the proposed algorithm, the program was coded in Python, run on Spyder 4.1.4, performed on a Notebook on Windows 10 with an Intel Core i7-8650U CPU at 1.90 GHz and 2.11 GHz with 16 GB RAM.

The experimental results are listed in Tables 9 and 10, illustrated in Figures 2−3, and analyzed in the following subsections. all 64 tests were and We generate all of the ($d$, $T$-MP$_{de}$s using the proposed algorithm in Stage 1. Then, the reliability of this MQPP$_{de}$ can be calculated in terms of all of the ($d$, $T$)-MP$_{de}$s obtained in Stage 1, as shown in Stage 2.

until it is

, , for UGFM and defined as

$$\pi_{i-1:J}\, z^J \otimes p_{i:I}\, z^I = \begin{cases} \pi_{i-1:J}\, p_{i:I}\, z^{(J \cup I) - \{i\}}, & \text{if } i \in J \\ \pi_{i-1:J}\, p_{i:I}\, z^J, & \text{if } i \notin J \end{cases} \tag{}$$

**STEP B1.** If sc < et $L_i(j(i)) = [L(j(i)-1) - \{i\}] \cup S_{i,j(i)}$ and $\tau_i = \tau_i - S_{i,j(i)}$.

**STEP B2.** If $L_i(j(i)) \neq \varnothing$, go to STEP B03. Otherwise, go to STEP B02.

**STEP B03.** If $\tau_i \subseteq S_{i,j(i)}$, $X$ is a feasible state vector and go to STEP B02.

**STEP B3.** If $j(i) < 2^{|\text{Deg}(i)|}$, let $x = x - 1$ and go to STEP B4.

**STEP B4.** If $x = 0$, then halt; otherwise, go to STEP B2.

**STEP B6.** Select the node, say $i$, with the smallest label in $L$, let $j(i) = 0$ and go to STEP B1.

**STEP B01.** If $\tau' = \varnothing$, one of the feasible state vector and let $R = R + \text{Pr}(X)$.the node with the smallest label in $L$, let such node be $i$, $j_i = 0$, and $L = L - \{i\}$.

**STEP B0.** Let the node label $i = 1$, the coordinate index $x_i = 1$, the node state label $j(i) = 1$, $L_i(j(i)) = S_{i,j(i)}$, the current reliability $R = 0$, $T_l = \{\, i \,\}$, $T_{-1} = \varnothing$, and go to STEP N2.

**STEP B01.** Let $L_i(j(i)) = L^*$. select the node, say $v_x$, with the smallest label in $L^*$, let $j(i) = 0$ and $L_i(0) = L^*$.

**STEP B1.** Let $L_i(j(i)) = [L(j(i)-1) - \{i\}] \cup S_{i,j(i)}$ and $\tau_i = \tau_i - S_{i,j(i)}$.

**STEP B2.** If $L_i(j(i)) \neq \varnothing$, go to STEP B03. Otherwise, go to STEP B02.

**STEP B02.** If $j(i) \leq 2^{|\mathrm{Deg}(i)|}$, let $j(i) = j(i) + 1$ and go to STEP B1. Otherwise, let $x = x - 1$ and go to STEP B4.

**STEP B03.** If $\tau_i \subseteq S_{i,j(i)}$, $X$ is a feasible state vector and go to STEP B02.

**STEP B3.** If $j(i) < 2^{|\mathrm{Deg}(i)|}$, let $x = x - 1$ and go to STEP B4.

**STEP B4.** If $x = 0$, then halt; otherwise, go to STEP B2.

**STEP B6.** Select the node, say $i$, with the smallest label in $L$, let $j(i) = 0$ and go to STEP B1.

**STEP B01.** If $\tau^* = \varnothing$, one of the feasible state vector and let $R = R + \Pr(X)$. the node with the smallest label in $L$, let such node be $i$, $j_i = 0$, and $L = L - \{i\}$.

; otherwise, let such node be $i$ and ,

; otherwise, let such node be $i$ and ,

**STEP B1.** Select the node with the smallest label in $L$. If there is no such node, let $j = j + 1$ and go to STEP B1; otherwise, let such node be $i$ and ,

**STEP B1.** Select the node, say $i$, with the smallest label in $L$., let such node be $i$, $j_i = 0$, and $L = L - \{i\}$.

**STEP N2.** Let $j_i = j_i + 1$ and $T^* = \{j \mid j \in [V(i) - T_i]\}$.

**STEP N3.** If $T^* = \varnothing$, go to STEP N6.

**STEP N4.** Let $T_l = T_{l-1} \cup T^*$, $X(j) = 0$ for all $j \in T^*$, and $P_l = P_{(l-1)} \times \Pr(S_{X(i)}(i))$.

**STEP N5.** If $|S_l| \geq N_{neighbor}$, let $R = R + P_l$ and go to STEP N2.

If node $i$ reaches its the maximum-state i.e., $j = 2^{|Deg(i)|} - 1$, go to N8.

**STEP N6.** Let new node $i$ be the node right after the current node $i$ in $T_l$ and go to STEP N3.

**STEP N7.** If there is no such node in STEP N6, go to STEP N2.

**STEP N8.** Let $l = l - 1$, $i$ be the node right before the current node $i$ in $T_l$, and go to STEP N1. If there is no such node $i$, halt and $\Pr(s, N_{area}) = R$ is the final probability that the wildfire can be spread out at least $N_{area}$ areas.

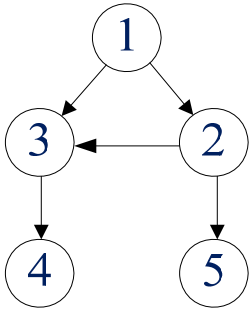

The number of coordinates in the state vectors may be different in the proposed BAT and the nodes connected may be present in the same state vectors.

For example, the initial state vector is $X = (1/3)$ if node 3, is the first identified area with a wildfire. The initial state vector always includes the first burned node and its state label starts from 1, as shown in STEP N0. Then, $X = (1/3, 0/5)$ due to $S_1(3) = \{5\}$ and node 3, does not reach its the maximum-state. Furthermore, the other areas that are not the first identified area must begin with state 0, as shown in STEP N4.

If $N_{area} = 2$, we obtain a possible state vector because the wildfire in node 3, is propagated to a neighbor area, i.e., node 5, by STEP N5. If $N_{area} > 2$, we should increase the state of node 5 by one because the current state of node 5 is zero, i.e., it does not reach another node. The aforementioned analysis is based on STEP N2 and is similar to that of the conventional BAT.

| $l$ | $i$ | $V(i)$ | Label[1] | Label[2] | $X(i)$ | $T^{\#}$ | $T_l = T_{l-1} \cup S^{\#}$ | $X$ | Probability[*] |
| --- | --- | --- | --- | --- | --- | --- | --- | --- | --- |
| 0 | 3 | {7, 6, 5} | 1 | 001 | 1 | {5} | {3, 5} | (1/3) | $p_{3,\{5\}}$ |
| 1 | 5 | {6, 4, 3, 2, 1, 0} | 0 | 000000 | 0 | ∅ | {3, 5} | (1/3, 0/5) | $p_{3,\{5\}}p_{5,\varnothing}$ |
| 2 | 5 | {6, 4, 3, 2, 1, 0} | 1 | 000001 | 1 | {5, 7} | {3, 5, 0, 7} | (1/3, 1/5) | $p_{3,\{5\}}p_{5,\{0\}}$ |
| 3 | 0 | {7, 5} | 0 | 00 | 0 | ∅ | {3, 5, 0, 7} | (1/3, 1/5, 0/0) | $p_{3,\{5\}}p_{5,\{0\}}p_{0,\varnothing}$ |
| 4 | 7 | {6, 4, 3, 2, 1, 0} | 0 | 000000 | 0 | ∅ | {3, 5, 0, 7} | (1/3, 1/5, 0/0, 0/7) | $p_{3,\{5\}}p_{5,\{0\}}p_{0,\{7\}}p_{7,\varnothing}$ |
| 5 | 7 | {6, 4, 3, 2, 1, 0} | 1 | 000001 | 1 | ∅ | {3, 5, 0, 7} | (1/3, 1/5, 0/0, 1/7) | $p_{3,\{5\}}p_{5,\{0\}}p_{0,\{7\}}p_{7,\{1\}}$ |

[1] decimal labels
[2] binary state labels
* $p_{i,I}$ = Pr($S_I(i)$) = Pr($X(i)$), where node subset $I$ is the $j$th state of node $I$ and $X(i) = j$.

## 4.1 DELETE

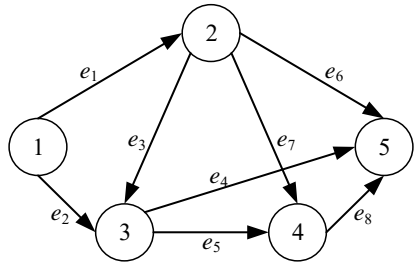

A state label is called a decimal state label or binary state label if it is represented by a decimal number or binary number, respectively. To recognize the state represented by a state label without losing any states, the notation used in BAT is adapted here to represent the state and state label of a node, i.e., $i \in V$, based on the following rules:

1. the number of digits is equal to Deg($i$) in the binary state label;

2. the nodes in the states of $i$ are arranged in the decreasing order of their node labels in $V(i)$;

3. the $i$th digit in the binary state label is equal to 0 or 1 if the $i$th node is included or excluded in the state, respectively;

**Table 5.** The first eight combinations of node $i$ based on $V(i)$ and Rules 1-3.

| Node \ State | 0 | 1 | 2 | 3 | 4 | 5 | 6 | 7 |
|---|---|---|---|---|---|---|---|---|
| 0 | ∅ | {5} | {7} | {5, 7} | | | | |
| 1 | ∅ | {5} | {6} | {7} | {5, 6} | {5, 7} | {6, 7} | {5, 6, 7} |
| 2 | ∅ | {5} | {6} | {6, 7} | | | | |
| 3 | ∅ | {5} | {6} | {7} | {5, 6} | {5, 7} | {6, 7} | {5, 6, 7} |
| 4 | ∅ | {5} | {6} | {7} | {5, 6} | {5, 7} | {6, 7} | {5, 6, 7} |
| 5 | ∅ | {0} | {1} | {2} | {0, 1} | {0, 2} | {1, 2} | {0, 1, 2} |
| 6 | ∅ | {1} | {2} | {3} | {1, 2} | {1, 3} | {2, 3} | {1, 2, 3} |
| 7 | ∅ | {0} | {1} | {3} | {0, 1} | {0, 3} | {1, 3} | {0, 1, 3} |

The state label is a decimal state label if there is no further clarification. For example, in Fig. 1, the degree of node 0 is 2, i.e., Deg(0) = 2, and its adjacent node subset is {7, 5}, i.e., $V(0)$ = {7, 5} (listed in the order of decreasing node labels by Rule 2). From Rules 1 and 3, these states ∅, {5}, {7}, and {5, 7} can be represented in the binary numbers 00, 01, 10, and 11, respectively, where the first/second digit in the binary numbers represents the state of node 7/5 based on the node listed order in $V(i)$. The corresponding decimal numbers of these binary numbers are state labels, i.e., state labels 0, 1, 2, and 3, are node subsets ∅, {5}, {7}, and {5, 7} because 0 = 00, 1 = 01, 2 = 10, and 3, = 11, respectively.

## 4.1 General results

The probability of one burned area, i.e., $N_{area}$ = 1, is 100% because there is always an ignition area in the wildfire propagation, e.g., Pr($i$, $N_{area}$=1) = 100% for all node $i$, as presented in Table 9.

From Table 9, it can be seen that the probability is decreased, number of state vectors is increased, and runtime is increased with an increased $N_{area}$. The aforementioned observation satisfies the real-life phenomena. Moreover, the number of state vectors and runtime are exponential for $N_{area}$ because the degree distribution in scale-free networks follows the power law.

## 4.2 Pr($i$, $N_{area}$) and Deg($i$)

Furthermore, from Table 9, if the adjacent node subsets of two different nodes are the same, e.g., $V(i) = V(j)$, we have Pr($i$, $N_{area}$) = Pr($j$, $N_{area}$) for all $N_{area}$. For example, for nodes 1, 3, and 4 in Fig. 1, $V(1) = V(3) = V(4)$, and their probabilities of having at least $N_{area}$ = 1, 2, …, 8 burned areas are identical: 1.0000000000, 0.9641535574, 0.9620059754, 0.9586257553, 0.9524496834, 0.9349168701, 0.8529978662, and 0.5144439304, respectively, i.e., Pr($i$, $N_{area}$) = Pr($j$, $N_{area}$) for $i$, $j$ = 1, 3, 4, and all values of $N_{area}$.

However, the total number of possible state vectors are different for a particular number of $N_{area}$. For example, for nodes 1, 3, and 4, $V(1) = V(3) = V(4)$, the number of vectors for nodes 1, 3, and 4 are different for 5 ≤ $N_{area}$, e.g., the 36276, 36212, and 36180 correspond to nodes 1, 3, and 4 when $N_{area}$ = 5.

It can be seen from Fig. 2, the higher degree node seems to have a higher probability in lower $N_{area}$ but lower probability in higher $N_{area}$. For example, nodes 5 and 6 exhibit a degree of 6, which is the highest among all the nodes in Fig. 1. $Pr(i = 6, N_{area})$ is always the highest among all other nodes for $N_{area}$ = 1, 2, …, 7 areas, but $Pr(i = 6, N_{area} = 8)$ ranks third from last. Similarly, $Pr(i = 5, N_{area})$ is less than $Pr(i = 6, N_{area})$ and its $Pr(i = 5, N_{area} = 8)$ is the least. On the contrary, $Pr(i = 2, N_{area})$ is always the least for $N_{area}$ = 1, 2, …, 7 but that of $Pr(i = 2, N_{area} = 8)$ is not the least.

Hence, it is necessary to extinguish wildfires in the areas with a higher degree in the beginning. However, the nodes with a lower degree for larger $N_{area}$ in the final stage of wildfire propagation require further attention.

### 4.3, $Pr(i, N_{area})$ and the Maximum-state PageRank

Moreover, the nodes with better ranking in PageRank show better ranking in the probability of smaller $N_{area}$; in contrast, these nodes with better ranking in 1 − (their PageRank values), i.e., the least PageRank values, show better ranking in the probability of larger $N_{area}$. The is because the higher degree of nodes always show higher values of PageRank and better $Pr(i, N_{area})$ for smaller values of $N_{area}$.

For example, in Table 10 and Fig. 3, A, B, …, H denote nodes 0, 1, …, 7 in Fig. 1, node 6 exhibits the best maximum-state PageRank and its values of $Pr(6, N_{area})$ correspond well for $N_{area}$ = 1, 2, …, 7; node 1 exhibits the least maximum-state PageRank values and its $Pr(0, N_{area})$ correspond well for $N_{area}$ = 8.

Hence, it is recommended to protect the areas with better maximum-state PageRank values at the beginning of the wildfire propagation and those with the least maximum-state PageRank values later. The observations further confirm the conclusion in Section 4.2.

**Table 10.** Node ranks based on Table 4 and PageRank values.

| $i$ \ $N_{area}$ | 1 | 2 | 3 | 4 | 5 | 6 | 7 | 8 | PR | 1-PR |
|---|---|---|---|---|---|---|---|---|---|---|
| 0 | 1 | 7 | 7 | 7 | 7 | 7 | 7 | **1** | 8 | **1** |
| 1 | 1 | 4 | 4 | 4 | 4 | 4 | 2 | 3 | 4 | 3 |
| 2 | 1 | 8 | 8 | 8 | 8 | 8 | 8 | 2 | 7 | 1 |
| 3 | 1 | 4 | 4 | 4 | 4 | 4 | 2 | 3 | 4 | 3 |
| 4 | 1 | 4 | 4 | 4 | 4 | 4 | 2 | 3 | 4 | 3 |
| 5 | 1 | 2 | 2 | 2 | 2 | 3 | 6 | 8 | 2 | 7 |
| 6 | **1** | **1** | **1** | **1** | **1** | **1** | **1** | 6 | **1** | 8 |

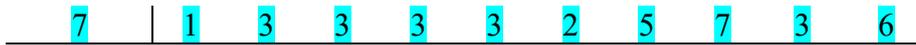

**Figure 3.** Node ranks based on maximum-state PageRank values.

## 4.4 Maximum-State PageRank

The proposed new BAT can calculate the correct value of $Pr(i, N_{area})$ for all nodes $i$ and $N_{area}$, i.e., the probability of the number of wildfire propagation areas, as presented in Table 9. However, the node degree distribution follows the power law, and the computation burden is increased following the power law based on the size of the scale-free network and value of $N_{area}$. The aforementioned obstacle may overcome the requirement for a high-performance computer or GPU for a larger-size scale-free network.

The aforementioned analysis shows that the number of wildfire propagation areas is highly dependent on the proposed concept: maximum-state PageRank is the summation of all PageRank values in the maximum state of the related node, described in Section 4.2 and 4.3.

Hence, in the aforementioned procedure, the computation burden may be avoided if we wish to identify the area that requires protection to reduce the wildfire propagation, i.e., we can consider $PR_{max}(i)$ if $Pr(i, N_{area})$ is not required.

**3.4 State PageRank, State Probability, and Maximum-state PageRank**

In the Barabási–Albert model [18] discussed in Section 2.2, the preferential attachment in STEP 2A emphasizes that the probability of a node obtaining a connection is proportional to its node degree. Hence, here, we define the occurrence spread probability of possible states, called the state probability.

After obtaining the PageRank values [30] of all nodes using the algorithm presented in Section 2.3, and the possible states of each node provided in Sections 3.1–3.3, the next step is to calculate the state PageRank and state probability.

The state PageRank and state probability represent the PageRank value and occurrence probability of the related state, respectively. We assume that $\mathrm{PR}(S_k(i))$ is the state PageRank of state $S_k(i)$ and $\mathrm{Pr}(S_k(i))$ is the probability to obtain $S_k(i)$, where node $i \in V$ and $k = 0, 1, \ldots, 2^{|\mathrm{Deg}(i)|}-1$. $\mathrm{PR}(S_k(i))$ is calculated based on the PageRank of all nodes in the corresponding state of node $i$ in terms of the probability of a node obtaining a connection proportional to its node degree, i.e.,

$$\mathrm{PR}(S_k(i)) = \sum_{|S_j(i)|=1 \text{ and } S_j(i) \subseteq S_k(i)} \mathrm{PR}(S_j(i)). \tag{9}$$

Because the summation of the probability of all possible states is equal to one, we can normalize the related PageRank values as shown in the following equation:

$$\mathrm{Pr}(S_k(i)) = \frac{\mathrm{PR}(S_k(i))}{\sum_{i=0}^{2^{|\mathrm{Deg}(i)|}-1} \mathrm{PR}(S_k(i))} \tag{10}$$

where

$$\mathrm{PR}(S_0(v)) = 0.5 \times \mathrm{Min}\{\, W(S_k(i)) \mid \text{for } |S_k(i)| = 1 \text{ and } k = 0, 1, \ldots, 2^{|\mathrm{Deg}(i)|}-1\}. \tag{11}$$

From the preferential attachment in the Barabási–Albert model [30], a newly added node is free to connect to any node in the network; however, the probability of connecting to a degree-two node is twice that of connecting to a degree-one node. Hence, we have Eqs. (9)–(11).